\documentclass[preprints,article,accept,moreauthors,pdftex]{Definitions/mdpi} 
\usepackage{changepage}
\usepackage{comment}
\firstpage{1} 
\pubvolume{1}
\issuenum{1}
\articlenumber{0}
\pubyear{2021}
\copyrightyear{2020}
\datereceived{} 
\dateaccepted{} 
\datepublished{} 
\hreflink{https://doi.org/} 

\usepackage{graphicx}
\usepackage{threeparttable}
\usepackage[normalem]{ulem}

\newcommand{\change}[2]{#2}

\newcommand{\changethird}[2]{#2}
\newcommand{\add}[1]{#1}
\newcommand{\addsec}[1]{#1}
\newcommand{\addthird}[1]{#1}
\newcommand{\remove}[1]{#1}

\Title{Deep Learning based Cardiac MRI segmentation: Do we need experts?}

\TitleCitation{Deep Learning based Cardiac MRI segmentation: Do we need experts?}

\Author{Youssef Skandarani $^{1,2, *}$, Pierre-Marc Jodoin $^{3}$ and Alain Lalande $^{1,4}$}

\AuthorNames{Youssef Skandarani, Pierre-Marc Jodoin and Alain Lalande}

\AuthorCitation{Skandarani, Y.; Jodoin, P-M.; Lalande, A.}

\address{%
$^{1}$ \quad University of Bourgogne Franche-Comte, Dijon, France\\
$^{2}$ \quad CASIS inc., Quetigny, France\\
$^{3}$ \quad Department of Computer Science, University of Sherbrooke, Sherbrooke, Canada\\
$^{4}$ \quad University Hospital of Dijon, Dijon, France}

\corres{Correspondence: youssef\_skandarani@etu.u-bourgogne.fr;}

\abstract{Deep learning methods are the de-facto solutions to a multitude of medical image analysis tasks. Cardiac MRI segmentation is one such application which, like many others, requires a large number of annotated data so a trained network can generalize well.  Unfortunately, the process of having a large number of manually curated images by medical experts is both slow and utterly expensive.
In this paper, we set out to explore whether expert knowledge is a strict requirement for the creation of annotated datasets that machine learning can successfully train on.  To do so, we gauged the performance of \changethird{a gold-standard neural network, the U-Net,}{three segmentation models, namely U-Net, Attention U-Net, and ENet,} trained with different loss functions on expert and non-expert groundtruth for cardiac cine-MRI segmentation. Evaluation was done with classic segmentation metrics (Dice index and Hausdorff distance) as well as clinical measurements, such as the ventricular ejection fractions and the myocardial mass.
Results reveal that generalization performances of a \changethird{U-Net}{segmentation neural network} trained on non-expert groundtruth data is, to all practical purposes, as good as on expert groundtruth data, in particular when the non-expert gets a decent level of training\add{, highlighting an opportunity for the efficient and cheap creation of annotations for cardiac datasets.}}


\begin{document}

\section*{Introduction}
Deep neural networks (and more specifically {\em convolutional neural networks}) have deeply percolated through healthcare R$\&$D addressing various problems such as survival prediction, disease diagnostics, image registration, anomaly detection, and segmentation of images be it Magnetic Resonance Images (MRI), Computed Tomography (CT) or ultrasound (US) to name a few~\cite{Litjens2017}.  The roaring success of deep learning methods is rightly attributed to the unprecedented amount of annotated data across domains.  But ironically, while solutions to decade-long medical problems are at hand~\cite{bernard2018deep}, the use of neural networks in day-to-day practice is still pending.  This can be explained in part by the following two observations.  First, while being accurate {\em on average}, neural networks can nonetheless be sometimes wrong~\cite{Painchaud20} as they provide no strict clinical guarantees.  In other words, any neural network within the intra-expert variability is excellent {\em on average} but not immune to sparse erroneous (yet degenerated) results which is problematic in clinical practice \cite{bernard2018deep}.  Second, machine learning methods are known to suffer from domain adaptation problems, one of the most glaring medical imaging issue of our times~\cite{Venkataramani19}.  As such, clinically accurate machine learning methods trained on a specific set of data almost always see their performances drop when tested on a dataset acquired following a different protocol.  These problems derive in good part from the fact that current datasets are still relatively small. According to Maier-Hein {\em et al.}~\cite{Maier18} most medical imaging challenges organized so far contain less than 100 training and testing cases.  This shows that medical applications cannot rely on  {\em very} large medical dataset encompassing tens of thousands of annotated data acquired in various conditions, with machines of various vendors showing clinical conditions and anatomical configurations of all kinds.

This is unlike non-medical computer vision problems which have had access for a long time to large and varied datasets such as ImageNet, Coco, PascalVOC, ADE20k, and Youtube-8M to name a few \cite{datasetlist}.  The annotation of these datasets rely on non-experts, often through online services like Mechanical Turk \cite{mturk}.  Unfortunately, obtaining similarly large annotated datasets in medical imaging is difficult.  The challenge stems from the nature of the data which is sensitive and requires navigating a complicated regulatory framework and privacy safeguards. Furthermore, labeling medical datasets is quite resource intensive and prohibitively costly as it requires a domain expertise.

For these reasons, the medical imaging literature have had an increasing number of publications whose goal is to compensate for the lack of expert annotations~\cite{karimi2020deep}.  While some methods leverage partly-annotated datasets~\cite{Can2018}, others use domain adaptation strategies to compensate for small training datasets~\cite{Choudhary20}.  Some other approaches artificially increase  the number of annotated data with Generative Adversarial Networks (GANs) \cite{goodfellow2014generative,skandarani2020effectiveness} and others use third-party neural networks to help experts annotate images more rapidly~\cite{girum2020fast}.

While these methods have been shown effective for their specific test cases, it is widely accepted that large manually-annotated datasets brings indisputable benefits~\cite{Sun17}.
In this work, \add{we depart from trying to improve the segmentation methods and focus on the datasets as} we challenge the idea that medical data\add{, cardiac cine MRI specifically,} \change{need}{needs} to be labeled by experts only and explore the consequences of using non-expert annotations on the generalization capabilities of a neural network. \add{Non-expert here refers to a non-physician who could not be regarded as a reference in the field.}  While non-expert annotations are easier and cheaper to get, they could be used to build larger datasets faster and at a reduced cost.

This idea was tested on cardiac cine-MRI segmentation. To this end, we had two non-experts labeling  cardiac cine-MRI images and compared the performance of neural networks trained on non-expert and expert data.
The evaluation between both approaches was done with geometric metrics (Dice index and Hausdorff distance) as well as clinical parameters namely, the ejection fraction for the left and right ventricles and the myocardial mass.

\section*{Methods and Data}
As mentioned before, medical data annotation requires a rightful expertise so the labeling can be used with full confidence.  Expert annotators are typically medical doctors or medical specialists whose training and experience are reliable sources of truth for the problem to solve.  These experts often have close collaborators  working daily with medical data, typically computer scientists, technicians, biophysicist, etc.  While their understanding of the data is real, these non-experts are typically  not considered as a reliable source of truth.  Non-expert are thus considered as people who can manually label data but whose annotations are biased and/or noisy and thus unreliable. 

In this study, two non-experts were asked to label 1902 cardiac images.  We defined a non-expert as someone with no professional expertise on cardiac anatomy nor cine-MR images. The Non-Expert 1 is a technician in biotechnology who received a 30 minute training by a medical expert on how to recognize and outline cardiac structures. \add{The training was done on a few examples where the expert showed what the regions of interest in the image look like and where their boundaries lie. Training also came with an introduction to the cardiac anatomy and its temporal dynamics.} The Non-Expert 2 is a computer scientist with 4 years of active research in cardiac cine-MRI with several months of training. \add{In the case of the Non-Expert 2, the training span several months where directions about the imaging modality as well as the anatomy and pathologies where thoroughly explained. In addition, fine delineation guidelines were provided to disambiguate good from poor annotations.} In this study, we both gauge the effect of training a neural network on non-expert data and also verify how the level of training of the non-experts impact the overall results.

 The non-experts were asked to delineate three structures of the heart namely, the left ventricular cavity (endocardial border), the left ventricle myocardium (epicardial border) and the endocardial border of the right ventricle.
 No further quality control was done to validate the non-expert annotations. Segmentations were used as-is for the subsequent tasks.

We used the gold standard for medical image segmentation U-Net \cite{ronneberger2015u} as the baseline network. \addthird{In addition, the well-known Attention U-Net \cite{Oktay2018attention} and ENet \cite{paszke2016enet} networks were trained in order to ensure that the results are affected by the differences in annotations and not the network architecture.} We first trained the \changethird{U-Net}{the segmentation models (U-Net, Attention U-Net and ENet)} on the original ACDC dataset (Automated Cardiac Diagnosis Challenge) ~\cite{bernard2018deep} with its associated groundtruth using a classical supervised training scheme, with a combined cross-entropy and Dice loss: 
\begin{equation}
\centering
L_{ce}= - \frac{1}{N} \sum_{i = 1}^{N} \sum_{k=1}^{3} y_{ki} \log{\hat{y}_{ki}}
\end{equation}
\begin{equation}
\centering
L_{dice} =  1 - \frac{1}{N} \bigg[ \sum_{i=1}^{N} \frac{2 \times \sum_{k} \hat{y}_{ki} y_{ki}}{\sum_{k}^{} \hat{y}_{ki} + \sum_{k}^{} y_{ki}  } \bigg]
\end{equation}
where $\hat{y}_{ki}$ is the probabilistic output for image $i \in N$ ($N$ is the number of images in the batch) and class $k \in \{1,2,3\}$ ($3$ is the number of classes). $\hat{y}$ is the predicted output of the network, $y$ is a one-hot encoding of the ground truth segmentation map.

We then re-trained the \changethird{U-Net}{neural networks} with the non-expert labels using the same training configuration.  Furthermore, considering that the non-expert annotations can be seen as noisy versions of the true annotation (i.e. $y' = y + \epsilon$ where $y'$ is the non-expert annotation, $y$ is the groundtruth and $\epsilon$ a random variable), we also trained the \changethird{network}{networks} with a  mean absolute error loss which, as shown by Ghosh et al~\cite{ghosh2017robust}, has the solve property of compensating for labeling inaccuracies.

\section*{Experimental Setup}
\begin{figure}[tp]
\centering
\label{fig_annot_compare}
\includegraphics[width=0.8\linewidth]{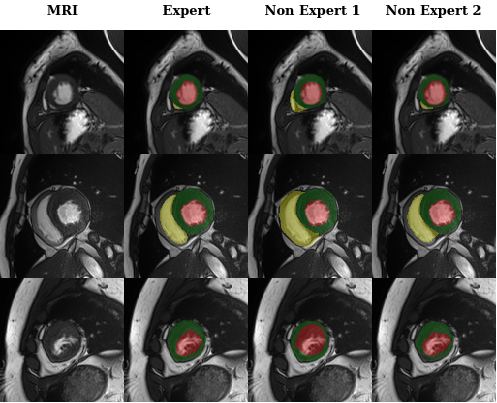}
  \caption{ACDC Annotation differences between Expert, Non Expert 1 and Non Expert 2.}
\end{figure}
To test whether non-expert annotated datasets hold any value for cardiac MRI segmentation, the following two cardiac cine MRI datasets were used:

\begin{itemize}
  \item Automated Cardiac Diagnosis Challenge (ACDC) dataset \cite{bernard2018deep}.  This dataset comprises 150 exams acquired at the University Hospital of Dijon (all from different patients).  It is divided into 5 evenly distributed subgroups (4 pathological plus 1 healthy subject groups) and split into 100 exams for training and 50 held out set for testing. \change{The exams were acquired using two MRI scanners of a magnetic strengths of 1.5T and 3T.}{The exams were acquired using two MRI scanners with different magnetic strengths (1.5T and 3T).} The pixel spacing varies from $0.7 mm$ to $1.9 mm$ with a slice spacing varying between $5 mm$ to $10 mm$. An example of images with the different expert and non-expert annotations is shown in Figure 1.
  \item Multi-Centre, Multi-Vendor \& Multi-Disease Cardiac Image Segmentation (M\&M) dataset \cite{campello2020}. This dataset consists of 375 cases from 3 different countries (Spain, Germany and Canada) totaling 6 different centres with 4 different MRI manufacturers (Siemens, General Electric, Philips and Canon). The cohort is composed of patients with hypertrophic and dilated cardiomyopathies as well as healthy subjects. The cine MR images were annotated by experienced clinicians from the respective centres. 
\end{itemize}

We trained the \change{U-Net}{segmentation models} on the 100 ACDC training subjects on either the expert and non-expert groundtruth data. Training was done with a fixed set of hyperparameters\addsec{, chosen through a cross-validated hyper-parameters search to best fit the 3 annotators,} without tuning it further.  The networks were trained three times in order to reduce the effect of the stochastic nature of the training process on the results.

As mentioned before, we first trained the neural networks on non-expert data with exactly the same setup as for the expert annotations.  Then, we retrained from scratch \changethird{the neural network }{the neural networks (U-Net, Attention U-Net and ENet)} with a L1 loss which was shown to be robust to noisy labels~\cite{ghosh2017robust}. 

We then tested in turn on the 50 ACDC test subjects and the 150 M\&Ms training data. The M\&Ms dataset constitutes data with groundtruth that is not biased towards either of the annotators of the training set, be it the expert or the non-expert. Moreover, testing on different datasets provides an inter-expert variability range as well as a domain generalization setup.


\begin{figure}[tp]
\centering
\label{fig_bad_pred}
\includegraphics[width=0.6\linewidth]{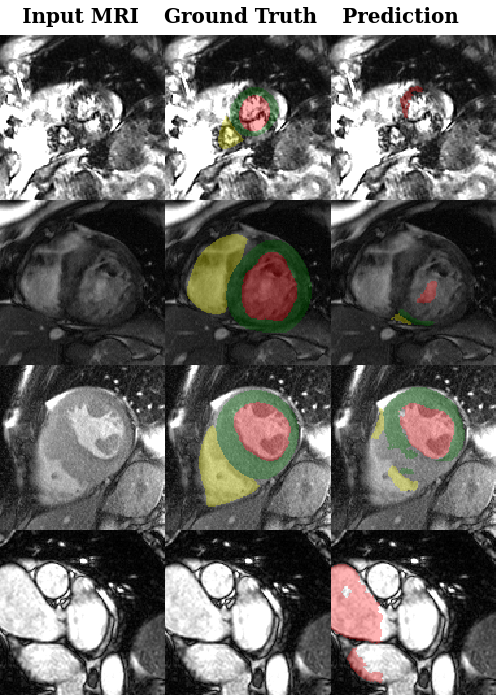}
  \caption{Examples of bad output segmentation on the M\&Ms dataset for the left ventricle, right ventricle and myocardium.\vspace{-0.5cm}}
\end{figure}

\section*{Results and Discussion}

The first set of results are laid out in Table 1.  It corresponds to standard geometrical metrics, i.e. the Dice score and the Hausdorff distance (HD) for the left ventricular (LV) cavity (Table 1), the myocardium (MYO) (considering the endocardial and epicardial of the left ventricle) (Table 2) and the cavity of the right ventricle (RV) (Table 3).  It also contains \add{the end-diastolic volume (EDV) as well as} the ejection fraction (EF) for the LV and the RV and the myocardial mass error.  For all three tables, the \changethird{UNet}{networks (U-Net, Attention U-Net and ENet)} has been trained on the ACDC training set and tested on the ACDC testing set and the M\&Ms training set.  

Results for the ACDC testing set reveal that for the LV (Table 1) the networks trained on the non-expert annotations (Non-Expert 1 as well as Non-Expert 2) manage to achieve performances that are statistically indistinguishable from those of the expert.  And that is true regardless of the training loss (CE+Dice vs MAE+Dice) and the metric (Dice, HD, and EF). The only exception is for the EF error for Non-Expert 1 with loss function MAE+Dice.

The situation however is more fuzzy for the MYO and the RV.  In both cases, we can see that results for the Non-Expert 1 are almost always worse than that of the expert, especially for the CE+Dice loss.  For example, there is a Dice score drop of $12\%$ on the myocardium.  Also, the clinical results on the RV (Table 3) show a clear gap between the Non-Expert 1 and the other annotators.  However, we can appreciate how the MAE+Dice loss improves results for both non-experts.  Overall for the MYO and the RV, results for the Non-Expert 2 are very close (if not better) than that of the expert. This is obvious when considering the average myocardial mass error in Table 2. Although, one recurrent result from our experiments is the hit-and-miss performance of all the evaluated networks on the M\&Ms dataset, where in a number of cases the output segmentation was completely degenerated as shown in Figure 2. \addthird{Moreover, the difference in segmentation performance between all the annotators is similar regardless of the segmentation network (U-Net, Attention U-Net or ENet) used, although Attention U-Net shows the best performance overall which is to be expected given its larger capacity.}

\begin{figure}[tp]
\centering
\includegraphics[width=1\linewidth]{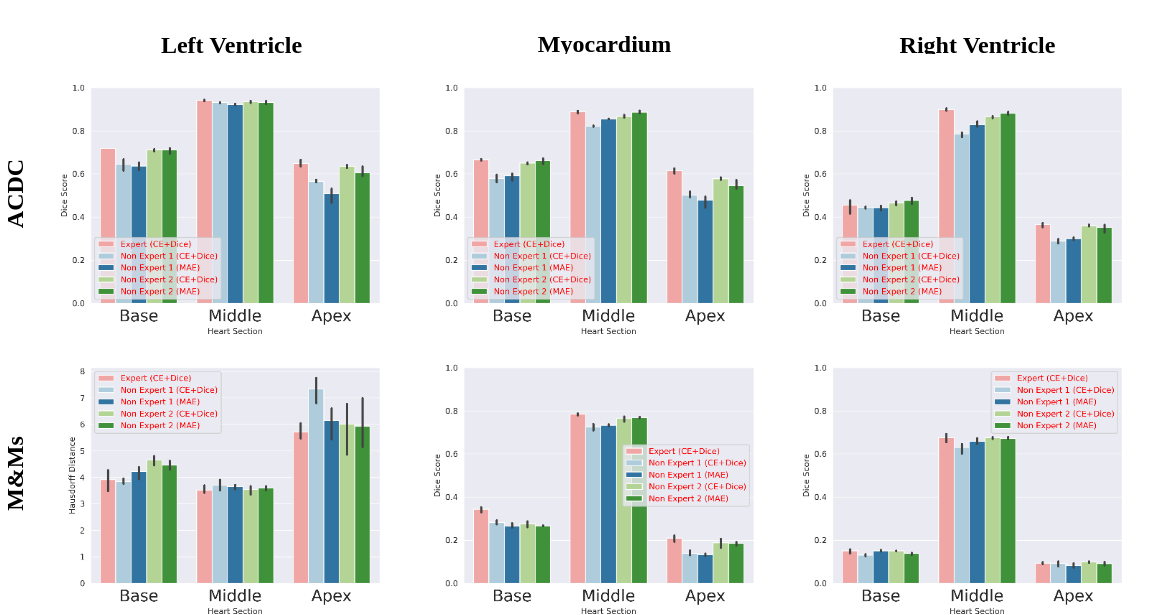}
  \caption{Dice score on ACDC and M\&Ms dataset per anatomical structure and per slice location.\vspace{-0.3cm}}
\label{fig_per_slice}
\end{figure}

\begin{figure}[tp]
\centering
\includegraphics[width=1\linewidth]{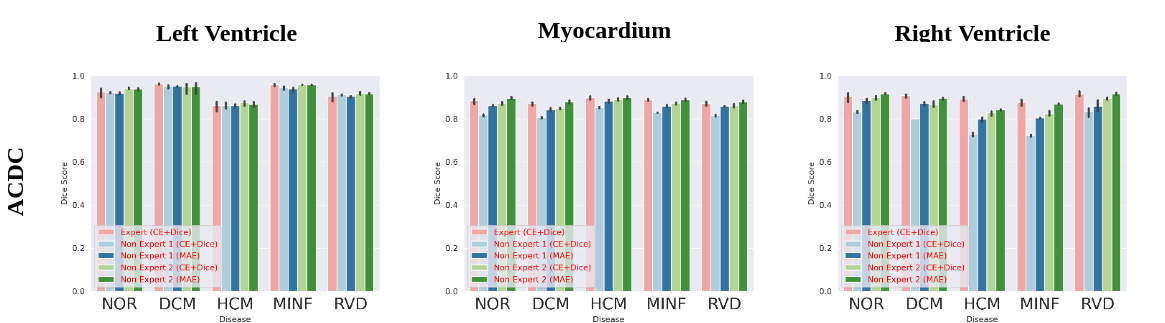}
  \caption{Dice score on ACDC per disease. NOR: Normal, DCM: Dilated Cardiomyopathy, HCM: Hypertrophic Cardiomyopathy, MINF: Myocardial Infarction, RVD: Right Ventricle Disease.\vspace{-0.2cm}}
\label{fig_per_disease}
\end{figure}

\add{Further analysis of the segmentation performance on the different sections of the heart (Figure \ref{fig_per_slice}), namely the base, the middle and the apex, show that the differences between the non-experts and the expert annotations lie heavily on the two ends of the heart. The performance gap is more pronounced on the apex for the three anatomical structures. In parallel, when we look at the performance from the disease groups (Figure \ref{fig_per_disease}), we can distinguish a relative similarity in the Dice score between the different annotators and disease groups.}


\begin{table}[H]
\centering
\footnotesize
\begin{adjustwidth}{-2.9cm}{}
\caption{Dice Score (DSC), Hausdorff Distance (HD), \add{ average Ejection Fraction (EF) error, and End Diastolic Volume (EDV)} of \textbf{left ventricular cavity} for \addthird{U-Net, Attention U-Net and E-Net}  with expert and non-expert annotations with data augmentation.}

\begin{threeparttable}[t]
\textbf{U-Net}
      \begin{tabular}{c|ccccccc}
        \hline
        Test set & ACDC trainset &  Loss & Avg. DSC & Avg. HD & \add{Avg. EF (\%)} & Avg EF err. & \add{Avg. EDV (mL)}\\ \hline

        & \textbf{Expert} &  CE+Dice & $0.92\pm0.08$ & $11.84\pm11.43$ & $47.57\pm19.62$ & $3.59\pm3.17$ & $175.70\pm69.03$\vspace{0.1cm}\\

        ACDC & Non-Expert 1   & $\left \{ \begin{matrix} \mbox{CE+Dice }\\ \mbox{MAE+Dice}\end{matrix} \right.$   &$\begin{matrix} 0.92\pm0.08\\ 0.91\pm0.08\end{matrix}$ & $\begin{matrix} 12.15\pm10.63\\ 12.80\pm10.48\end{matrix}$ & $\begin{matrix} 50.27\pm20.00\\ 50.37\pm20.14\end{matrix}$ & $\begin{matrix} 4.37\pm3.69\\ 4.52^*\pm4.32\end{matrix}$ & $\begin{matrix} 173.05\pm67.06\\ 173.57\pm70.29\end{matrix}$ \vspace{0.1cm}\\

        & Non-Expert 2   & $\left \{ \begin{matrix} \mbox{CE+Dice }\\ \mbox{MAE+Dice}\end{matrix} \right.$   &$\begin{matrix} \boldsymbol{0.93\pm0.09}\\ 0.92\pm0.09\end{matrix}$ & $\begin{matrix} \boldsymbol{11.68\pm12.36}\\ 11.76\pm12.34\end{matrix}$ & $\begin{matrix} 49.73\pm20.70\\ 49.45\pm20.68\end{matrix}$ & $\begin{matrix}\boldsymbol{3.35\pm3.21}\\ 3.69\pm3.59\end{matrix}$ & $\begin{matrix} 182.98\pm72.51\\ 181.19\pm70.93\end{matrix}$ \vspace{0.1cm}\\ \hline
        
        & \textbf{Expert} &  CE+Dice & $0.86\pm0.11$ & $15.03\pm8.07$ & $54.22\pm15.11$ & $7.47\pm5.79$ & $162.57\pm60.22$ \vspace{0.1cm}\\

        M\&Ms & Non-Expert 1   & $\left \{ \begin{matrix} \mbox{CE+Dice }\\ \mbox{MAE+Dice}\end{matrix} \right.$   &$\begin{matrix} 0.86\pm0.10\\ 0.86\pm0.10\end{matrix}$ & $\begin{matrix} 11.76\pm6.04\\ \boldsymbol{11.52\pm5.55}\end{matrix}$ & $\begin{matrix} 56.81^*\pm15.37\\ 57.14^*\pm15.23\end{matrix}$ & $\begin{matrix} 5.53^*\pm5.22\\ \boldsymbol{5.39^*\pm5.38}\end{matrix}$ & $\begin{matrix} 149.75^*\pm57.33\\ 151.08^*\pm58.08\end{matrix}$  \vspace{0.1cm}\\

        & Non-Expert 2   & $\left \{ \begin{matrix} \mbox{CE+Dice }\\ \mbox{MAE+Dice}\end{matrix} \right.$   &$\begin{matrix} \boldsymbol{0.88\pm0.09}\\ 0.88\pm0.10\end{matrix}$ & $\begin{matrix} 11.58\pm6.93\\ 11.84\pm7.93\end{matrix}$ & $\begin{matrix} 56.70^*\pm15.33\\ 56.74^*\pm15.45\end{matrix}$ & $\begin{matrix} 5.54^*\pm4.57\\ 6.17^*\pm5.11\end{matrix}$ & $\begin{matrix} 164.95\pm62.00\\ 165.78\pm62.00\end{matrix}$ \vspace{0.1cm}\\ \hline
      \end{tabular}
      
    \textbf{Attention U-Net}
      \begin{tabular}{c|ccccccc}
        \hline
        Test set & ACDC trainset &  Loss & Avg. DSC & Avg. HD & Avg. EF (\%) & Avg. EF err & Avg. EDV \\ \hline

        & \textbf{Expert} &  CE+Dice & $0.92\pm0.08$ & $11.78\pm11.14$ &  $46.83\pm19.50$ & $3.81\pm3.63$ & $177.77\pm71.08$ \vspace{0.1cm}\\

        ACDC & Non-Expert 1   & $\left \{ \begin{matrix} \mbox{CE+Dice }\\ \mbox{MAE+Dice}\end{matrix} \right.$   &$\begin{matrix} 0.92\pm0.08\\ 0.92\pm0.09\end{matrix}$ & $\begin{matrix} 12.07\pm11.20\\ 13.42\pm12.62\end{matrix}$ &  $\begin{matrix} 50.27\pm20.07\\ 50.35\pm20.28\end{matrix}$  & $\begin{matrix} 4.28\pm3.94\\ 4.74^*\pm4.15\end{matrix}$ & $\begin{matrix} 170.86\pm67.86\\ 173.43\pm67.61\end{matrix}$  \vspace{0.1cm}\\

        & Non-Expert 2   & $\left \{ \begin{matrix} \mbox{CE+Dice }\\ \mbox{MAE+Dice}\end{matrix} \right.$   &$\begin{matrix} \boldsymbol{0.93\pm0.07}\\ 0.93\pm0.08\end{matrix}$ & $\begin{matrix} \boldsymbol{11.05\pm11.45}\\ 12.52\pm11.62\end{matrix}$ & $\begin{matrix} 50.56\pm21.24\\ 49.78\pm21.10\end{matrix}$ & $\begin{matrix} \boldsymbol{3.50\pm3.86}\\ 3.93\pm3.96\end{matrix}$ & $\begin{matrix} 180.28\pm70.50\\ 178.82\pm68.16\end{matrix}$ \vspace{0.1cm}\\ \hline
        
        & \textbf{Expert} &  CE+Dice & $\boldsymbol{0.88\pm0.09}$ & $11.90\pm7.38$ & $55.77\pm15.23$ & $\boldsymbol{5.59\pm4.73}$ & $164.93\pm62.67$ \vspace{0.1cm}\\

        M\&Ms & Non-Expert 1   & $\left \{ \begin{matrix} \mbox{CE+Dice }\\ \mbox{MAE+Dice}\end{matrix} \right.$   &$\begin{matrix} 0.86\pm0.12\\ 0.86\pm0.10\end{matrix}$ & $\begin{matrix} 11.90\pm7.73\\ 12.40\pm6.79\end{matrix}$ & $\begin{matrix} 57.67\pm17.34\\ 57.40\pm17.44\end{matrix}$ & $\begin{matrix} 6.64^*\pm6.45\\ 6.23\pm6.18\end{matrix}$ & $\begin{matrix} 148.25^*\pm57.84\\ 148.22^*\pm57.37\end{matrix}$  \vspace{0.1cm}\\

        & Non-Expert 2   & $\left \{ \begin{matrix} \mbox{CE+Dice }\\ \mbox{MAE+Dice}\end{matrix} \right.$   &$\begin{matrix} 0.87\pm0.10\\ 0.87\pm0.09\end{matrix}$ & $\begin{matrix} \boldsymbol{11.10\pm6.59}\\ 11.14\pm6.43\end{matrix}$ & $\begin{matrix} 58.51^*\pm16.33\\ 58.20^*\pm16.34\end{matrix}$ & $\begin{matrix} 5.99\pm5.01\\ 6.29^*\pm5.36\end{matrix}$ & $\begin{matrix} 161.20\pm61.40\\ 161.29\pm62.34\end{matrix}$ \vspace{0.1cm}\\ \hline
      \end{tabular}
      \textbf{E-Net}
      \begin{tabular}{c|ccccccc}
        \hline
        Test set & ACDC trainset &  Loss & Avg. DSC & Avg. HD & Avg. EF (\%) & Avg. EF err & Avg. EDV \\ \hline

        & \textbf{Expert} &  CE+Dice & $0.92\pm0.09$ & $12.28\pm11.38$ & $46.68\pm19.06$ & $3.94\pm3.86$ &  $178.48\pm70.68$ \vspace{0.1cm}\\

        ACDC & Non-Expert 1   & $\left \{ \begin{matrix} \mbox{CE+Dice }\\ \mbox{MAE+Dice}\end{matrix} \right.$   &$\begin{matrix} 0.91\pm0.09\\ 0.91\pm0.10\end{matrix}$ & $\begin{matrix} 12.72\pm12.28\\ 13.07\pm15.00\end{matrix}$ & $\begin{matrix} 48.64\pm20.33\\ 50.10\pm19.99\end{matrix}$ & $\begin{matrix} 4.31\pm4.99\\ 4.15\pm3.65\end{matrix}$ &  $\begin{matrix} 172.60\pm67.69\\ 171.53\pm67.29\end{matrix}$  \vspace{0.1cm}\\

        & Non-Expert 2   & $\left \{ \begin{matrix} \mbox{CE+Dice }\\ \mbox{MAE+Dice}\end{matrix} \right.$   &$\begin{matrix} 0.92\pm0.09\\ \boldsymbol{0.92\pm0.08}\end{matrix}$ & $\begin{matrix} 12.18\pm14.01\\ \boldsymbol{11.69\pm11.16}\end{matrix}$ & $\begin{matrix} 49.05\pm20.52\\ 48.83\pm20.57\end{matrix}$ & $\begin{matrix} \boldsymbol{3.25\pm3.09}\\ 3.45\pm3.53\end{matrix}$ &  $\begin{matrix} 180.77\pm71.20\\ 181.07\pm70.45\end{matrix}$ \vspace{0.1cm}\\ \hline
        
        & \textbf{Expert} &  CE+Dice & $0.86\pm0.10$ & $14.46\pm7.54$ & $53.91\pm16.04$ &  $7.19\pm5.98$ & $165.77\pm61.59$ \vspace{0.1cm}\\

        M\&Ms & Non-Expert 1   & $\left \{ \begin{matrix} \mbox{CE+Dice }\\ \mbox{MAE+Dice}\end{matrix} \right.$   &$\begin{matrix} 0.86\pm0.12\\ 0.86\pm0.12\end{matrix}$ & $\begin{matrix} 13.64\pm10.60\\ \boldsymbol{11.90\pm8.59}\end{matrix}$ & $\begin{matrix} 53.10\pm21.96\\ 55.90^{\dagger}\pm18.26\end{matrix}$ & $\begin{matrix} 7.41\pm12.67\\ \boldsymbol{5.81^{*\dagger}\pm7.64}\end{matrix}$ & $\begin{matrix} 152.59^*\pm59.87\\ 150.79^*\pm57.79\end{matrix}$  \vspace{0.1cm}\\

        & Non-Expert 2   & $\left \{ \begin{matrix} \mbox{CE+Dice }\\ \mbox{MAE+Dice}\end{matrix} \right.$   &$\begin{matrix} \boldsymbol{0.87\pm0.10}\\ \boldsymbol{0.87\pm0.10}\end{matrix}$ & $\begin{matrix} 12.39\pm7.63\\ 12.72\pm8.41\end{matrix}$ & $\begin{matrix} 53.53\pm17.49\\ 54.99\pm16.29\end{matrix}$ & $\begin{matrix} 6.39\pm7.25\\ 6.76\pm5.56\end{matrix}$ &  $\begin{matrix} 165.05\pm61.73\\ 165.36\pm60.71\end{matrix}$ \vspace{0.1cm}\\ \hline
      \end{tabular}
      \begin{tablenotes}
        \item[*] p-value $< 0.05$ non-expert vs expert.
        \item[$\dagger$] p-value $< 0.05$ MAE vs CE.
      \end{tablenotes}
      \end{threeparttable}
      \end{adjustwidth}
\end{table}

\begin{table}[H]
\footnotesize
\begin{adjustwidth}{-2cm}{}
\centering
\caption{Dice Score, Hausdorff Distance and Mass error for the \textbf{myocardium} for \addthird{U-Net, Attention U-Net and E-Net} with expert and non-expert annotations with data augmentation.}
\begin{threeparttable}[t]
\textbf{U-Net}
      \begin{tabular}{c|ccccc}
        \hline
        Test set & ACDC trainset &  Loss & Avg. DSC & Avg. HD & Avg. Mass err.\\ \hline

        & \textbf{Expert} &  CE+Dice & $0.88\pm0.03$ & $11.02\pm6.25$ & $16.88\pm9.66$ \vspace{0.1cm}\\

        ACDC & Non-Expert 1   & $\left \{ \begin{matrix} \mbox{CE+Dice }\\ \mbox{MAE+Dice}\end{matrix} \right.$   &$\begin{matrix} 0.82\pm0.03\\ 0.86\pm0.03\end{matrix}$ & $\begin{matrix} 11.45\pm4.88\\ 11.69\pm5.18\end{matrix}$ & $\begin{matrix} 36.40^*\pm13.37\\ 15.93^{\dagger}\pm8.95\end{matrix}$  \vspace{0.1cm}\\

        & Non-Expert 2   & $\left \{ \begin{matrix} \mbox{CE+Dice }\\ \mbox{MAE+Dice}\end{matrix} \right.$   &$\begin{matrix} 0.87\pm0.03\\ \boldsymbol{0.89\pm0.03}\end{matrix}$ & $\begin{matrix} \boldsymbol{10.13\pm5.02}\\ 10.43\pm5.00\end{matrix}$ & $\begin{matrix} 11.68^*\pm7.74\\ \boldsymbol{6.58^{*\dagger}\pm5.93}\end{matrix}$  \vspace{0.1cm}\\ \hline
        
        & \textbf{Expert} &  CE+Dice & $\boldsymbol{0.80\pm0.06}$ & $17.57\pm7.65$ & $17.36\pm12.87$ \vspace{0.1cm}\\

        M\&Ms & Non-Expert 1   & $\left \{ \begin{matrix} \mbox{CE+Dice }\\ \mbox{MAE+Dice}\end{matrix} \right.$   &$\begin{matrix} 0.76\pm0.08\\ 0.78\pm0.09\end{matrix}$ & $\begin{matrix} 14.91\pm8.60\\ 14.32\pm7.95\end{matrix}$ & $\begin{matrix} 22.14^{*}\pm19.71\\ 18.78^{\dagger}\pm17.80\end{matrix}$  \vspace{0.1cm}\\

        & Non-Expert 2   & $\left \{ \begin{matrix} \mbox{CE+Dice }\\ \mbox{MAE+Dice}\end{matrix} \right.$   &$\begin{matrix} 0.80\pm0.07\\ 0.80\pm0.07\end{matrix}$ & $\begin{matrix} 14.29\pm7.53\\ \boldsymbol{13.76\pm6.79}\end{matrix}$ & $\begin{matrix} \boldsymbol{16.36\pm13.69}\\ 19.44^{*\dagger}\pm16.79\end{matrix}$  \vspace{0.1cm}\\ \hline
        
      \end{tabular}
      
 \textbf{Attention U-Net}
     \begin{tabular}{c|ccccc}
        \hline
        Test set & ACDC trainset &  Loss & Avg. DSC & Avg. HD & Avg. Mass err \\ \hline

        & \textbf{Expert} &  CE+Dice & $\boldsymbol{0.88\pm0.03}$ & $11.57\pm7.13$ & $17.95\pm10.28$ \vspace{0.1cm}\\

        ACDC & Non-Expert 1   & $\left \{ \begin{matrix} \mbox{CE+Dice }\\ \mbox{MAE+Dice}\end{matrix} \right.$   &$\begin{matrix} 0.84\pm0.03\\ 0.86\pm0.03\end{matrix}$ & $\begin{matrix} \boldsymbol{10.27\pm5.25}\\ 13.11\pm8.65\end{matrix}$ & $\begin{matrix} 33.99^*\pm13.17\\ 20.12^{\dagger}\pm8.82\end{matrix}$  \vspace{0.1cm}\\

        & Non-Expert 2   & $\left \{ \begin{matrix} \mbox{CE+Dice }\\ \mbox{MAE+Dice}\end{matrix} \right.$   &$\begin{matrix} 0.87\pm0.03\\ \boldsymbol{0.88\pm0.03}\end{matrix}$ & $\begin{matrix} 10.86\pm6.17\\ 11.53\pm6.62\end{matrix}$ & $\begin{matrix} 13.76^{*}\pm8.99\\ \boldsymbol{9.68^{*\dagger}\pm7.32}\end{matrix}$ \vspace{0.1cm}\\ \hline
        
        & \textbf{Expert} &  CE+Dice & $\boldsymbol{0.82\pm0.05}$ & $15.85\pm9.84$ & $16.33\pm12.75$ \vspace{0.1cm}\\

        M\&Ms & Non-Expert 1   & $\left \{ \begin{matrix} \mbox{CE+Dice }\\ \mbox{MAE+Dice}\end{matrix} \right.$   &$\begin{matrix} 0.78\pm0.08\\ 0.77\pm0.09\end{matrix}$ & $\begin{matrix} 13.95\pm8.04\\ 16.19\pm9.78\end{matrix}$ & $\begin{matrix} 19.44^{*}\pm17.61\\ 20.80^{*}\pm20.08\end{matrix}$  \vspace{0.1cm}\\

        & Non-Expert 2   & $\left \{ \begin{matrix} \mbox{CE+Dice }\\ \mbox{MAE+Dice}\end{matrix} \right.$   &$\begin{matrix} 0.80\pm0.07\\ 0.78\pm0.08\end{matrix}$ & $\begin{matrix} \boldsymbol{13.43\pm7.85}\\ 13.75\pm7.56\end{matrix}$ & $\begin{matrix} \boldsymbol{16.17\pm13.58}\\ 26.89^{*\dagger}\pm21.43\end{matrix}$ \vspace{0.1cm}\\ \hline
        
      \end{tabular}      
      
\textbf{E-Net}
      \begin{tabular}{c|ccccc}
        \hline
        Test set & ACDC trainset &  Loss & Avg. DSC & Avg. HD & Avg. Mass err \\ \hline

        & \textbf{Expert} &  CE+Dice & $\boldsymbol{0.87\pm0.03}$ & $10.99\pm5.61$ & $17.14\pm9.05$ \vspace{0.1cm}\\

        ACDC & Non-Expert 1   & $\left \{ \begin{matrix} \mbox{CE+Dice }\\ \mbox{MAE+Dice}\end{matrix} \right.$   &$\begin{matrix} 0.82\pm0.04\\ 0.86\pm0.04\end{matrix}$ & $\begin{matrix} 11.91\pm7.80\\ 11.51\pm6.03\end{matrix}$ & $\begin{matrix} 39.81^{*}\pm15.58\\ 21.26^{*\dagger}\pm9.49\end{matrix}$ \vspace{0.1cm}\\

        & Non-Expert 2   & $\left \{ \begin{matrix} \mbox{CE+Dice }\\ \mbox{MAE+Dice}\end{matrix} \right.$   &$\begin{matrix} 0.85\pm0.04\\ \boldsymbol{0.87\pm0.03}\end{matrix}$ & $\begin{matrix} 10.96\pm6.11\\ \boldsymbol{10.49\pm5.16}\end{matrix}$ & $\begin{matrix} 15.64\pm9.59\\ \boldsymbol{7.83^{*\dagger}\pm6.24}\end{matrix}$ \vspace{0.1cm}\\ \hline
        
        & \textbf{Expert} &  CE+Dice & $0.80\pm0.07$ & $15.81\pm7.55$ & $17.86\pm15.63$ \vspace{0.1cm}\\

        M\&Ms & Non-Expert 1   & $\left \{ \begin{matrix} \mbox{CE+Dice }\\ \mbox{MAE+Dice}\end{matrix} \right.$   &$\begin{matrix} 0.77\pm0.10\\ 0.78\pm0.09\end{matrix}$ & $\begin{matrix} 16.12\pm12.23\\ \boldsymbol{14.51\pm9.26}\end{matrix}$ & $\begin{matrix} 28.40^{*}\pm23.91\\ 18.35^{\dagger}\pm17.10\end{matrix}$  \vspace{0.1cm}\\

        & Non-Expert 2   & $\left \{ \begin{matrix} \mbox{CE+Dice }\\ \mbox{MAE+Dice}\end{matrix} \right.$   &$\begin{matrix} \boldsymbol{0.80\pm0.06}\\ 0.79\pm0.08\end{matrix}$ & $\begin{matrix} 15.18\pm8.41\\ 15.69\pm8.39\end{matrix}$ & $\begin{matrix} \boldsymbol{16.08\pm14.98}\\ 18.14^{\dagger}\pm15.82\end{matrix}$  \vspace{0.1cm}\\ \hline
        
      \end{tabular}      
      
      \begin{tablenotes}
        \item[*] p-value $< 0.05$ non-expert vs expert.
        \item[$\dagger$] p-value $< 0.05$ MAE vs CE.
      \end{tablenotes}
     \end{threeparttable}
     \end{adjustwidth}
\end{table}

\begin{table}[H]
\footnotesize
\begin{adjustwidth}{-2.4cm}{}

\centering
\caption{Dice Score, Hausdorff Distance, \add{average Ejection Fraction (EF) error, and End Diastolic Volume (EDV)} for the \textbf{right ventricular cavity} for \addthird{U-Net, Attention U-Net and E-Net} with expert and non-expert annotations with data augmentation
.}
    \begin{threeparttable}[t]
    \textbf{U-Net}
      \begin{tabular}{c|ccccccc}
        \hline
        Test set & ACDC trainset &  Loss & Avg. DSC & Avg. HD & \add{Avg. EF (\%)} & Avg. EF err. & \add{Avg. EDV (mL)} \\ \hline

        & \textbf{Expert} &  CE+Dice & $\boldsymbol{0.90\pm0.06}$ & $14.72\pm6.12$ & $42.77\pm15.25$ & $\boldsymbol{6.69\pm6.21}$ & $183.73\pm70.68$\vspace{0.1cm}\\

        ACDC & Non-Expert 1   & $\left \{ \begin{matrix} \mbox{CE+Dice }\\ \mbox{MAE+Dice}\end{matrix} \right.$   &$\begin{matrix} 0.78\pm0.11\\ 0.84\pm0.09\end{matrix}$ & $\begin{matrix} 19.59\pm8.07\\ 16.78\pm7.23\end{matrix}$ & $\begin{matrix} 32.22^*\pm13.49\\ 35.51^{*\dagger}\pm14.64\end{matrix}$ & $\begin{matrix} 14.77^*\pm9.88\\ 13.52^*\pm8.88\end{matrix}$ & $\begin{matrix} 216.16^*\pm69.96\\ 181.61^{\dagger}\pm61.85\end{matrix}$ \vspace{0.1cm}\\

        & Non-Expert 2   & $\left \{ \begin{matrix} \mbox{CE+Dice }\\ \mbox{MAE+Dice}\end{matrix} \right.$   &$\begin{matrix} 0.86\pm0.07\\ 0.89\pm0.07\end{matrix}$ & $\begin{matrix} 15.25\pm6.00\\ \boldsymbol{14.65\pm6.34}\end{matrix}$ & $\begin{matrix} 40.10\pm14.42\\ 41.61\pm14.85\end{matrix}$ & $\begin{matrix} 7.98\pm7.34\\ 7.38\pm7.27\end{matrix}$ & $\begin{matrix} 200.67^*\pm73.35\\ 186.59\pm71.56\end{matrix}$ \vspace{0.1cm}\\ \hline
        
        & \textbf{Expert} &  CE+Dice & $\boldsymbol{0.82\pm0.15}$ & $\boldsymbol{16.91\pm12.65}$ & $52.90\pm19.79$ & $\boldsymbol{9.12\pm12.73}$ & $145.92\pm61.94$ \vspace{0.1cm}\\

        M\&Ms & Non-Expert 1   & $\left \{ \begin{matrix} \mbox{CE+Dice }\\ \mbox{MAE+Dice}\end{matrix} \right.$ &$\begin{matrix} 0.74\pm0.15\\ 0.78\pm0.15\end{matrix}$ & $\begin{matrix} 22.51\pm15.82\\ 16.93\pm10.95\end{matrix}$ & $\begin{matrix} 37.68^*\pm21.06\\ 38.75^*\pm36.59\end{matrix}$ & $\begin{matrix} 17.88^{*}\pm18.47\\ 18.12^{*}\pm33.08\end{matrix}$ & $\begin{matrix} 173.34^*\pm70.88\\ 144.36^{\dagger}\pm60.57\end{matrix}$ \vspace{0.1cm}\\

        & Non-Expert 2   & $\left \{ \begin{matrix} \mbox{CE+Dice }\\ \mbox{MAE+Dice}\end{matrix} \right.$   &$\begin{matrix} 0.79\pm0.16\\ 0.81\pm0.17\end{matrix}$ & $\begin{matrix} 20.31\pm17.56\\ 18.05\pm15.60\end{matrix}$ & $\begin{matrix} 44.11^*\pm29.89\\ 39.17^*\pm91.83\end{matrix}$ & $\begin{matrix} 12.86^{*}\pm26.88\\ 20.72^{*}\pm88.32\end{matrix}$ & $\begin{matrix} 168.76^*\pm66.89\\ 149.17^{\dagger}\pm66.41\end{matrix}$ \vspace{0.1cm}\\ \hline
      \end{tabular}
      
      \textbf{Attention U-Net}
            \begin{tabular}{c|ccccccc}
        \hline
        Test set & ACDC trainset &  Loss & Avg. DSC & Avg. HD & Avg. EF (\%) & Avg. EF err. & Avg. EDV (mL) \\ \hline

        & \textbf{Expert} &  CE+Dice & $\boldsymbol{0.89\pm0.07}$ & $16.62\pm7.71$ & $38.73\pm15.58$ & $8.71\pm7.46$ &  $194.05\pm71.01$ \vspace{0.1cm}\\

        ACDC & Non-Expert 1   & $\left \{ \begin{matrix} \mbox{CE+Dice }\\ \mbox{MAE+Dice}\end{matrix} \right.$ &$\begin{matrix} 0.79\pm0.10\\ 0.83\pm0.09\end{matrix}$ & $\begin{matrix} 20.60\pm9.19\\ 20.08\pm11.27\end{matrix}$ & $\begin{matrix} 30.12^*\pm13.11\\ 33.00^*\pm14.62\end{matrix}$ & $\begin{matrix} 15.96^*\pm9.27\\ 13.57^{*\dagger}\pm9.68\end{matrix}$ & $\begin{matrix} 218.32^*\pm69.08\\ 198.36^{\dagger}\pm68.13\end{matrix}$  \vspace{0.1cm}\\

        & Non-Expert 2   & $\left \{ \begin{matrix} \mbox{CE+Dice }\\ \mbox{MAE+Dice}\end{matrix} \right.$   &$\begin{matrix} 0.86\pm0.07\\ 0.88\pm0.09\end{matrix}$ & $\begin{matrix} 16.33\pm7.92\\ \boldsymbol{16.30\pm9.18}\end{matrix}$ & $\begin{matrix} 37.96\pm15.19\\ 39.65\pm15.58\end{matrix}$ & $\begin{matrix} 9.20\pm7.15\\ \boldsymbol{8.31\pm8.28}\end{matrix}$ &  $\begin{matrix} 200.98\pm71.70\\ 185.30\pm70.51\end{matrix}$ \vspace{0.1cm}\\ \hline
        
        & \textbf{Expert} &  CE+Dice & $\boldsymbol{0.82\pm0.15}$ & $\boldsymbol{17.94\pm12.54}$ & $50.23\pm15.46$ & $\boldsymbol{9.39\pm8.21}$ & $156.45\pm62.88$ \vspace{0.1cm}\\

        M\&Ms & Non-Expert 1   & $\left \{ \begin{matrix} \mbox{CE+Dice }\\ \mbox{MAE+Dice}\end{matrix} \right.$ &$\begin{matrix} 0.76\pm0.13\\ 0.77\pm0.14\end{matrix}$ & $\begin{matrix} 20.96\pm11.07\\ 24.24\pm14.61\end{matrix}$ & $\begin{matrix} 37.69^*\pm19.13\\ 37.94^*\pm18.12\end{matrix}$ & $\begin{matrix} 17.86^*\pm15.87\\ 17.67^*\pm15.47\end{matrix}$ & $\begin{matrix} 179.85^*\pm67.54\\ 167.74^{*\dagger}\pm60.49\end{matrix}$  \vspace{0.1cm}\\

        & Non-Expert 2   & $\left \{ \begin{matrix} \mbox{CE+Dice }\\ \mbox{MAE+Dice}\end{matrix} \right.$   &$\begin{matrix} 0.79\pm0.15\\ 0.80\pm0.17\end{matrix}$ & $\begin{matrix} 18.09\pm13.02\\ 20.72\pm15.26\end{matrix}$ & $\begin{matrix} 47.28^*\pm19.16\\ 44.72^*\pm27.57\end{matrix}$ & $\begin{matrix} 12.20^*\pm12.84\\ 13.22^*\pm23.62\end{matrix}$ & $\begin{matrix} 159.95\pm64.95\\ 151.40^{\dagger}\pm63.41\end{matrix}$ \vspace{0.1cm}\\ \hline
      \end{tabular}
      
      \textbf{E-Net}
            \begin{tabular}{c|ccccccc}
        \hline
        Test set & ACDC trainset &  Loss & Avg. DSC & Avg. HD & Avg. EF (\%) & Avg. EF err. & Avg. EDV (mL) \\ \hline

        & \textbf{Expert} &  CE+Dice & $\boldsymbol{0.88\pm0.07}$ & $\boldsymbol{15.76\pm6.38}$ & $38.65\pm15.27$ & $\boldsymbol{9.11\pm7.17}$ & $192.65\pm70.02$ \vspace{0.1cm}\\

        ACDC & Non-Expert 1   & $\left \{ \begin{matrix} \mbox{CE+Dice }\\ \mbox{MAE+Dice}\end{matrix} \right.$   &$\begin{matrix} 0.77\pm0.11\\ 0.83\pm0.10\end{matrix}$ & $\begin{matrix} 22.14\pm9.01\\ 17.47\pm8.29\end{matrix}$ & $\begin{matrix} 30.12^*\pm13.12\\ 34.48^{*\dagger}\pm13.68\end{matrix}$ & $\begin{matrix} 16.29^*\pm10.02\\ 13.00^{*\dagger}\pm9.11\end{matrix}$ & $\begin{matrix} 223.07^{*}\pm71.29\\ 189.00^{\dagger}\pm65.27\end{matrix}$  \vspace{0.1cm}\\

        & Non-Expert 2   & $\left \{ \begin{matrix} \mbox{CE+Dice }\\ \mbox{MAE+Dice}\end{matrix} \right.$   &$\begin{matrix} 0.84\pm0.09\\ 0.87\pm0.08\end{matrix}$ & $\begin{matrix} 16.19\pm6.29\\ 16.46\pm7.34\end{matrix}$ & $\begin{matrix} 36.25\pm14.13\\ 36.87\pm13.84\end{matrix}$ & $\begin{matrix} 10.46\pm7.91\\ 9.90\pm7.91\end{matrix}$ & $\begin{matrix} 203.49\pm71.15\\ 194.54\pm69.42\end{matrix}$ \vspace{0.1cm}\\ \hline
        
        & \textbf{Expert} &  CE+Dice & $0.79\pm0.18$ & $19.57\pm16.32$ & $48.86\pm20.66$ & $\boldsymbol{12.06\pm13.44}$ &  $151.92\pm66.12$ \vspace{0.1cm}\\

        M\&Ms & Non-Expert 1   & $\left \{ \begin{matrix} \mbox{CE+Dice }\\ \mbox{MAE+Dice}\end{matrix} \right.$   &$\begin{matrix} 0.72\pm0.16\\ 0.78\pm0.14\end{matrix}$ & $\begin{matrix} 28.38\pm17.94\\ 19.30\pm12.58\end{matrix}$ & $\begin{matrix} 34.20^{*}\pm30.51\\ 41.63^{*\dagger}\pm19.67\end{matrix}$ & $\begin{matrix} 22.27^{*}\pm26.54\\ 14.83^{*\dagger}\pm15.06\end{matrix}$ & $\begin{matrix} 178.67^*\pm72.98\\ 156.65^{\dagger}\pm61.99\end{matrix}$  \vspace{0.1cm}\\

        & Non-Expert 2   & $\left \{ \begin{matrix} \mbox{CE+Dice }\\ \mbox{MAE+Dice}\end{matrix} \right.$   &$\begin{matrix} 0.79\pm0.14\\ \boldsymbol{0.80\pm0.14}\end{matrix}$ & $\begin{matrix} \boldsymbol{19.13\pm12.09}\\ 20.73\pm13.26\end{matrix}$ & $\begin{matrix} 41.88^{*}\pm18.65\\ 42.94^{*}\pm16.39\end{matrix}$ & $\begin{matrix} 14.17^*\pm13.61\\ 13.35\pm11.36\end{matrix}$ & $\begin{matrix} 166.83^*\pm66.72\\ 169.43^*\pm63.54\end{matrix}$ \vspace{0.1cm}\\ \hline
      \end{tabular}

      \begin{tablenotes}
        \item[*] p-value $< 0.05$ non-expert vs expert.
        \item[$\dagger$] p-value $< 0.05$ MAE vs CE.
      \end{tablenotes}
     \end{threeparttable}
     \end{adjustwidth}
\end{table}

%
Our experiments also reveal some interesting results on the M\&Ms dataset, a dataset with different acquisition settings than \remove{for} ACDC. In that case, we see the gap in performance between the expert and the non-expert decrease substantially.  For example, when comparing the Non-Expert 1 results with MAE+Dice loss and those from the expert annotation, we see that the Dice difference for the RV went from a 6\%  on the ACDC dataset to a mere 4\% on the M\&Ms dataset. But overall, there again, results by Non-Expert 2 are similar (and sometimes better) than that of the expert.

Through out our experiments, the performances of \changethird{the networks}{the three neural networks (U-Net, Attention U-Net and ENet)} trained on the Non-Expert 2 annotations with MAE+Dice loss has been roughly on par, if not better, with those trained on the expert annotations.  This is especially true for the LV.  For the Non-Expert 1, most likely due to a lack of proper training,  results on both test sets and most MYO and RV metrics are worst than that of the expert.  In fact, a statistical test reveal that results from the Non-Expert 1 are almost always statistically different than that of the expert. We also evaluated the statistical difference between the CE+Dice and the MAE+Dice losses and observed that the MAE+Dice loss provides overall better results for both non experts.

Overall on M\&Ms, while the expert got a better MYO mass error and a better RV EF error, the MYO HD is lower for Non-Expert 2 and the Dice score and the RV HD of Non-Expert 2 are statistically similar. These results underline the idea that well-trained non-expert and expert annotations could be used interchangeably to build reliable annotated cardiac datasets. \addsec{In contrast, the number of non-experts we evaluated might be considered a limitation of our study, however, }\add{ this still provides encouraging results for settings where experts are not readily available to annotate whole datasets, but could provide training to a non-expert to effectively annotate in their stead. We leave the investigation on more datasets to future works that could transpose the setup to more difficult problems and a larger number of non-experts.}
\addthird{Our work comes to supplement previous endavours that rely on non-experts to annotate medical datasets, Heim {\em et~al.} \cite{heim2018large} showcased the ability of crowdsourced expertise to reliably annotate liver dataset although their approach proposes initial segmentations to the non-expert which might bias their decision. Likewise, Ganz {\em et~al.} \cite{ganz2016crowd} proposed to make use of non-experts as a crowdsourced error detection framework. In contrast, our approach evaluated the effectiveness of non-expert knowledge without any prior input. This further reinforces the idea that crowdsourced medical annotation are a viable solution to the lack of data.}
\section*{Conclusion}
In this work, we studied the usefulness of training deep learning models with non-expert annotations for the segmentation of cardiac MR images. \add{The need for medical experts was probed in a comparative study with non-physician sourced labels.} Through framing the problem of relying on non-expert annotations as noisy data, we managed to obtain good performance on two public datasets,  one of which was used to emulate an out-of-distribution dataset. We found that training a deep neural network\addthird{, regardless of its capacity (U-Net, Attention U-Net or ENet),} with data labeled by a well-trained non-expert achieves comparable performance than on expert data. Moreover, the performance gap between the networks with non-expert and expert annotations on the out-of-distribution dataset was less pronounced  than the  gap on the training dataset. Future endeavors could focus on crowd sourcing large-scale medical datasets and tailoring approaches that take their noisiness into account.



\vspace{6pt} 



\authorcontributions{Y.S: writing manuscript, developing software, experiment design, performing experiment. \\
    P-M.J : initial idea, writing manuscript, experiment design, data analysis. \\
    A.L: initial idea, writing manuscript, resource management, data analysis.}

\funding{This research received no external funding}

\acknowledgments{We would like to acknowledge Ivan Porcherot for the tremendous work he did annotating the datasets.}

\conflictsofinterest{The authors declare no conflict of interest.} 

\end{paracol}
\reftitle{References}

\end{document}